\documentclass[prb,twocolumn,showpacs,amsmath,amssymb]{revtex4-1}

\usepackage{graphicx}
\usepackage{bm}
\usepackage{makeidx}

\newcommand{\bmnin}{\mbox{$\beta$--Mn$_{1-x}$In$_x$~}}
\newcommand{\bmnal}{\mbox{$\beta$--Mn$_{1-x}$Al$_x$~}}
\newcommand{\bmn}{\mbox{$\beta$--Mn}}

\begin{document}

\title{Structural and dynamical study of moment localization in \bmnin}

\author{J. R. Stewart}
\email{ross.stewart@stfc.ac.uk}
\author{A. D. Hillier}
\affiliation{ISIS, STFC, Rutherford Appleton Laboratory, Didcot, OX11 0QX, UK}
\author{J. M. Hillier}
\affiliation{Oxford University Department of Education, 15 Norham Gardens, Oxford, OX2 6PY}
\altaffiliation{Institut Laue-Langevin, 6, rue Jules Horowitz, F-38042 Grenoble, France}
\author{R. Cywinski}
\affiliation{School of Applied Sciences, University of Huddersfield, Queensgate, Huddersfield, HD1 3DH, UK}

\date{\today}

\begin{abstract}
We have used neutron scattering and muon spin relaxation ($\mu$SR) to investigate the structural and magnetic properties of the $\beta$--phase of elemental manganese doped with dilute amounts of indium.  \bmn\ is an example of a topologically frustrated antiferromagnetically correlated metal - but which remains paramagnetic at all temperatures.  The addition of In to \bmn\ results in a vast volume expansion of the lattice, and would therefore be expected to have a major effect on the stability and localization of the Mn moment - as observed in, for example, Ru and Al doped \bmn\ alloys.  We find that In doping in \bmn\ results in a short-range ordered spin-glass like ground state, similar to that of Al-doped  \bmn\ but with residual low frequency spin fluctuations. This is in contrast to Ru doping which results in the stabilization of a long-range ordered Mn moment
\end{abstract}

\maketitle

\section{\label{intro}Introduction}
The magnetic phase diagrams and physical properties of nearly all the magnetically ordered 3d-elements are well-characterised and understood.  This is not the case for metallic manganese.  The low temperature--stable (up to 983 K) $\alpha$--phase of manganese exhibits a fiendishly complex bcc structure with no less than 58 Mn atoms in a single crystallographic unit cell distributed amongst four non--equivalent lattice positions.~\cite{Lawson94}  The magnetic structure of $\alpha$--Mn below 95 K is complex non--collinear antiferromagnetic, with 6 non-equivalent Mn moments of various sizes participating in the magnetic order.~\cite{Lawson94,Hobbs03}  The much simpler fcc $\gamma$--phase of Mn is another well-studied example of an itinerant electron antiferromagnet.  $\gamma$--Mn, which is only stable in a very limited temperature range, 1373 K to 1407 K, may nevertheless be maintained to much lower temperatures by the addition of small amounts of transition metal impurities such as Cu, Fe, Ni and Pd.   All these dilute alloys are antiferromagnetic at high temperatures, with the {\it extrapolated} N\'eel temperature for pure $\gamma$--Mn being given as 540 K.~\cite{Bacon57,Ling98}  In addition, both $\alpha$--Mn and $\gamma$--Mn show strong magneto-elastic coupling effects, with the antiferromagnetic transitions in each being accompanied by first order martensitic structural transitions from cubic to tetragonal symmetry.~\cite{Lawson94,Basinski54}

In contrast, the simple cubic $\beta$--phase of elemental Mn (stable from 983~K to 1373~K, between the $\alpha$-- and $\gamma$--phases) is an itinerant electron paramagnet at all temperatures.  The temperature dependences of the Mn spin-lattice relaxation rate,~\cite{Kohori93} $1/T_1 \propto \sqrt T$, and the low temperature electrical resistivity,~\cite{Stewart02} $\rho \propto T^{3/2}$, as well as a strongly enhanced T--linear (electronic) component of the specific heat capacity,~\cite{Shinkoda79} $\gamma \simeq 70$ mJ mol$^{-1}$K$^{-2}$, show that \bmn\ is a strongly exchange enhanced {\it nearly} antiferromagnetic metal, on the verge of moment formation, according to the self-consistent renormalization (SCR) theory of Moriya.~\cite{Moriya85}    However, the tuning of the exchange enhancement in \bmn\, either by increase of the d--electron density of states at the Fermi energy via doping by transition metal substituents ({\it e.g.} Fe~\cite{Vinogradova97} or Co~\cite{Stewart09b,Funahashi84} ) or by the application of chemical pressure brought about by the introduction of dilute non-magnetic impurities ({\it e.g.} Al~\cite{Nakamura97,Stewart08,Stewart99} or Sn~\cite{Nakai94,Nakai96}) rarely results in a long--range ordered magnetic ground state.  Only in the cases of Ru, Ir or Os doping does incommensurate antiferromagnetic long--range order appear;~\cite{Miyakawa05,Miyakawa03,Sasao01}  and in these cases, the magnetic transition temperatures and magnetic spin configurations are remarkably robust, with an incommensurate wavevector ($\bm{k} = (u u u)$ where $u=0.278$) and N\'eel temperatures ($T_N \simeq 130$ K) almost independent of impurity type, or doping level.~\cite{Miyakawa05,Stewart07b}   

The $\beta$--phase of manganese presents a simple cubic primitive cell with non--centrosymmetric space group P4$_1$32, containing 20 Mn atoms shared between two non--equivalent crystallographic sites at Wyckoff positions $8c$ (Mn1 site) and $12d$ (Mn2 site).  The Mn1 sublattice is extremely close packed with a near-neighbour distance of 2.36 \AA.  By contrast, the Mn2 sublattice is considerably more widely spaced with a near-neighbour distance of 2.66 \AA.  This observation has lead to the conjecture that the 3d--electron levels of Mn1 atoms are well below the Fermi level - such that the Mn1 contribution to the density of states at the Fermi level,~$\mathcal{D}(\epsilon_F)$, is small resulting in little or no magnetic moment for Mn1.~\cite{Nakamura97}  This observation is apparently backed up by Mn NMR~\cite{Kohori93} and NQR~\cite{Hama04} measurements, which show that the Mn spin-lattice relaxation rate is 20 times stronger for Mn2 atoms as for Mn1.  This picture is also supported by the observation of long--range magnetic order in $\beta$--Mn(Ru), $\beta$--Mn(Os) and $\beta$--Mn(Ir), but not in $\beta$--Mn(Al) or $\beta$--Mn(Sn).  Experimentally it is found that Ru, Os and Ir impurities all occupy the ``non--magnetic'' Mn1 site,~\cite{Sasao01,Yamauchi00,Miyakawa03b} while Al and Sn occupy the ``magnetic'' Mn2 site.~\cite{Stewart08,Nakai94}  Thus the conjecture is that Ru, Ir and Os dopants do not interfere with the magnetic Mn2 sublattice with a resulting long--range ordered ground state, but that the presence of non--magnetic Al and Sn on the Mn2 sublattice is disadvantageous to the formation of a long--range ordered magnetic configuration due to disruption of the Mn--Mn exchange pathways.~\cite{Stewart08,Nakamura97}  

\begin{figure}
\includegraphics[width=3.2in]{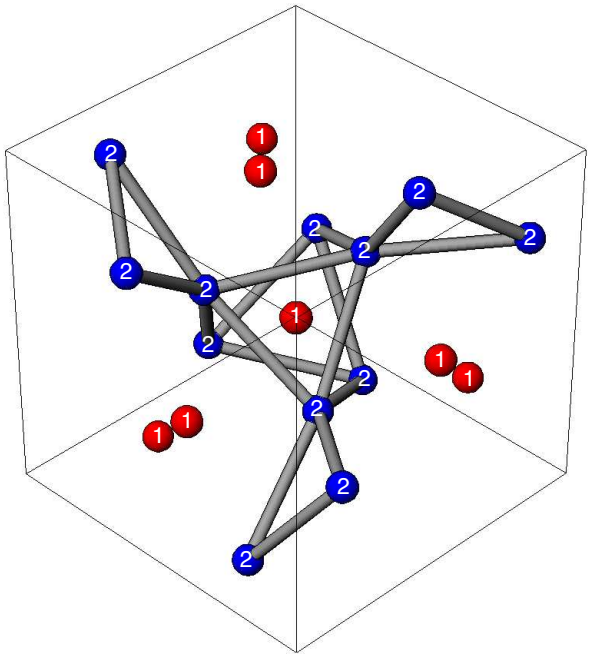}
\caption{\label{structure}(color online) The projection along the $[1~1~1]$ axis of the \bmn\ primitive cell.  The ``distorted windmill" of corner sharing triangles perpendicular to local $<1~1~1>$ axes connecting the Mn2 sublattice is depicted ~\cite{Kripyakevich60,Shoemaker78}. The Mn1 atoms are also shown, unconnected.}
\end{figure}

It has also been suggested that the lack of long--range magnetic order in pure \bmn\ and most \bmn\ alloys is due to the presence of geometrical frustration inherent in the \bmn\ lattice.  This issue was first addressed by the group of Shiga and co--workers ~\cite{Shiga94,Nakamura97} who pointed out the possibility of topological frustration of antiferromagnetically correlated moments in the Mn2 sublattice. They drew on the previous work of Shoemaker et.~al.~\cite{Shoemaker78} who described the Mn2 sublattice as a distorted network of corner-sharing triangles - termed a ``distorted windmill'', with the plane of each triangle being normal to each of the local $<1~1~1>$ axes, and each Mn2 atom shared between 3 triangles (see Fig.~\ref{structure}).  This structure is reminiscent of the well-studied frustrated hyperkagome system, Gd$_3$Ga$_5$O$_{12}$, which is a spin--liquid at low temperatures.~\cite{Petrenko98}
Assuming that the \bmn\ Mn2 sublattice is populated by Heisenberg spins with only first near-neighbour interactions, Canals and Lacroix~\cite{Canals00} have calculated the energy eigenvalues of the interaction matrix $J(q)$ and shown that the Mn2 sublattice is indeed frustrated under these conditions, with a dispersionless  and two-fold degenerate maximum eigenvalue.  They also show that further neighbour interactions relieve this frustration, resulting in either a \mbox{$q=0$} or incommensurate long--range antiferromagnetic spin--configuration, dependent on the sign of the second--neighbour interactions.  In this picture, the substitution of non--magnetic impurities such as Al and Sn on to the Mn2 sublattice, would be expected to (at least partially) relieve the frustration of the Mn2 lattice, and therefore promote magnetic order.   Static order is indeed observed in these systems~\cite{Stewart99,Nakai94} - but neutron scattering investigations have shown that spin--correlations persist only over a short--range, reminiscent of a spin--glass or cluster--glass type picture.~\cite{Stewart08}  While the topology of the Mn spin in \bmn\ is certain to have a major influence on the magnetic ground--state, it is also clear that a localized Heisenberg--spin picture with short--range interactions is likely not to be applicable to an itinerant paramagnet such as \bmn\ where all the indications are of delocalized spins (e.g. a flat temperature independent bulk susceptibility).~\cite{Nakamura97}

In an attempt to further investigate the promotion of static order in \bmn\ alloys we have undertaken a comprehensive study of the spin--configurational and dynamical properties of \bmn\ doped with indium.  Indium doping is interesting due to the fact that, according to the seminal survey of \bmn\ alloys by Kohara and Asayama,~\cite{Kohara74} the volume expansion due to chemical pressure of the \bmn\ lattice is strongest on the introduction of indium impurities, with a volume increase $\Delta V/V \simeq 1$\% per at.\% In.  Thus, tuning of the Mn--Mn near neighbour distances can be accomplished with minimal disruption of the \bmn\ coordination.  For example, to  achieve a 1.7 \% volume expansion - the point at which we have previously observed a spin--liquid to spin--glass transition in the $\beta$--MnAl system, a doping level of around 10 at.\% Al is needed, compared to only 1.6 at.\%. In.  As in the cases of Al and Sn, we show using neutron diffraction that In impurities predominantly reside on the magnetic Mn2 sublattice.  The magnetic ground--state spin--configurations, measured using neutron polarization analysis will be presented.  We also use muon spin relaxation ($\mu$SR) to show that In impurities promote damping of the Mn spin--fluctuations and formation of a static ground state at low temperatures.  There remain, however, strong Mn spin--fluctuations at low temperatures coexistent with the static magnetic component.

\section{\label{prep}Sample Preparation and Characterisation}
All of the \bmnin\ samples used in this study were prepared by the arc--melting technique in which high purity constituent metals are melted together using an electric arc in a low pressure pure argon atmosphere.   After melting, the \bmn\ phase was stabilized by annealing the melted ingots under a low argon pressure in sealed quartz containers at $\sim$~1170 K for 24 hours.  The \bmn\ phase was retained at room temperature by rapid quenching of the ingots in water.  Preliminary structural analysis of the samples, and verification of the $\beta$--phase of Mn (and more importantly, the absence of the antiferromagnetic $\alpha$--Mn phase) was achieved using x--ray diffraction.  We were successful in preparing \bmnin\ with In concentrations of $x=0$, 0.01, 0.02, 0.03, 0.04, 0.05 and 0.07.  An attempt to make a $x=0.1$ sample was found to be unsuccessful, with evidence of a mixed phase, and In insolubility. 

\begin{figure}
\includegraphics[width=3.2in]{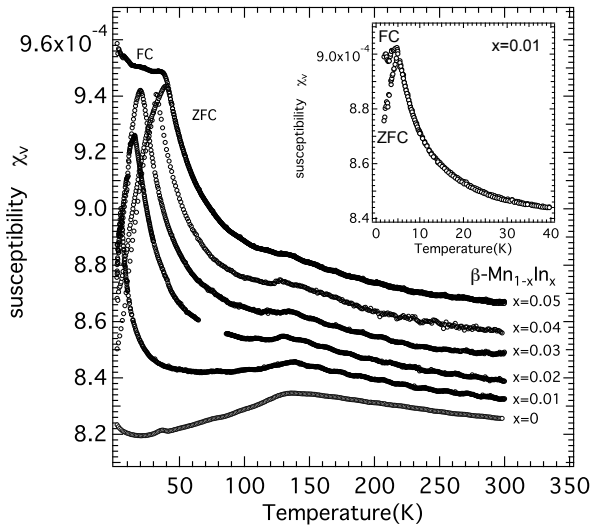}
\caption{\label{sus} The DC susceptibility of \bmnin\ samples with $x=$0, 0.01, 0.02, 0.03, 0.04 and 0.05, measured in a field of 10 mT.  The plots show the zero--field--cooled (ZFC) susceptibility of the samples, with the field--cooled (FC) susceptibility of the $x=0.05$ sample also shown.  The inset shows the low temperature detail of the ZFC and FC susceptibility of the $x=0.01$ sample.}
\end{figure}
\begin{figure}
\includegraphics[width=3.2in]{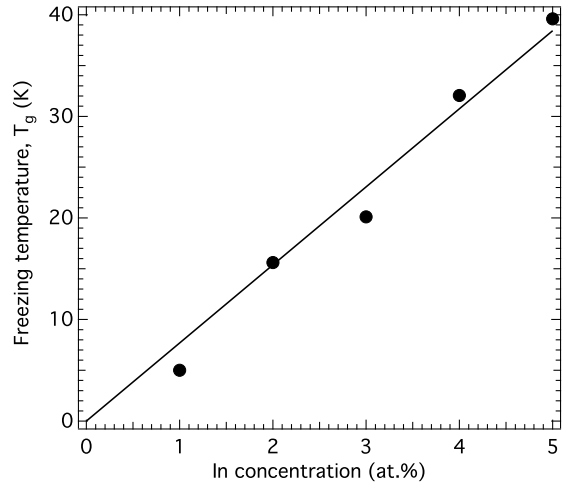}
\caption{\label{tg} The indium concentration dependence of the peak in the DC magnetic susceptibility (shown in Fig.~\ref{sus}), taken to indicate a spin--glass freezing transition temperature, $T_g$. The straight line highlights the roughly linear increase of $T_g$ on indium doping.}
\end{figure}
Magnetic characterization of the \bmnin\ samples was carried out using an Oxford Instruments Mk. I vibrating sample magnetometer (VSM) and later, a Quantum Design MPMS SQUID susceptometer.  Plots of the temperature dependence of the DC susceptibility - as measured using the SQUID susceptometer - are shown in Fig.~\ref{sus}.  For pure \bmn\ the DC susceptibility is flat as a function of temperature, indicating a response typical of an itinerant Pauli paramagnet.  In common with previous data taken by Nakamura and co--workers~\cite{Nakamura97} there is a shallow peak in the DC susceptibility for pure \bmn\ at $T \simeq 140$ K.  On indium doping of as little as 1 at. \%, the response becomes sharply peaked at finite temperature, with a characteristic bifurcation of the field--cooled (FC) and zero--field--cooled (ZFC) branches of the susceptibility.  Such history dependence is often taken to indicate a spin--glass freezing temperature, $T_g$.   The peak temperature of the ZFC susceptibility is plotted as a function of indium concentration in Fig.~\ref{tg}.  The increase in $T_g$ as a function of concentration is at a rate of approximately 7.7 K per at. \%\ In.   The development of a T-dependent bulk susceptibility and spin--glass freezing in \bmnin\ is akin to the response seen in \bmnal\ up to much greater Al concentrations, $x=0.3$, except that the freezing temperature increases much slower at a rate of around 1.7 K per at. \% Al.~\cite{Nakamura97, Stewart99}  Interestingly, the feature at 140 K observed in pure \bmn\ is present in all the samples, and apparently independent of indium concentration. 

\section{\label{diffraction}Neutron Powder Diffraction}
\begin{figure}
\includegraphics[width=3.2in]{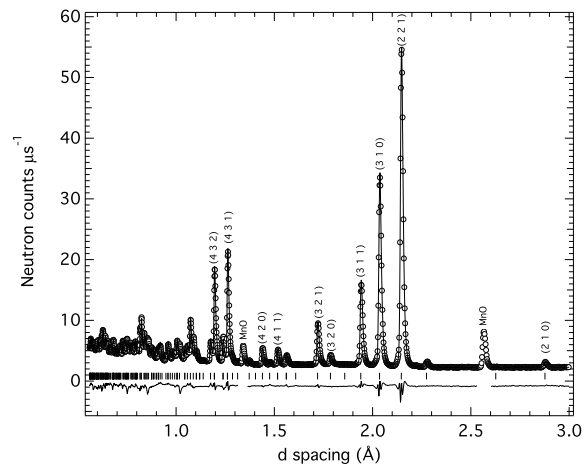}
\caption{\label{lad_diffraction}Time-of-flight neutron diffraction pattern of $\beta$--Mn$_{0.95}$In$_{0.05}$ taken on the LAD diffractometer at ISIS, UK.  The solid line is a model fit to the data, determined by Rietveld refinement using the {\footnotesize GSAS} package~\cite{Larson94}.  The difference curve between the data and the model is shown.}
\end{figure}
\begin{figure}
\includegraphics[width=3.2in]{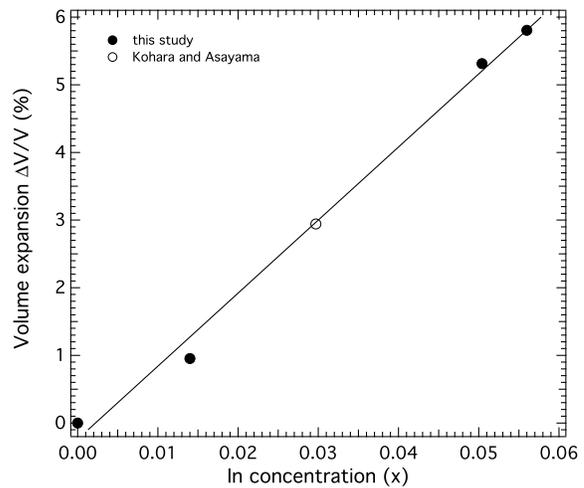}
\caption{\label{vol_exp}Volume expansion of \bmnin as a function of In concentration - data from this study are black circles, the white circle is taken from the work of Kohara~\cite{Kohara74}.  The solid line shows the rate of expansion found to be $1.08 \pm 0.05$ \% per at.\% In}
\end{figure}

Structural characterisation of \bmnin\ samples with nominal concentrations $x=0$, 0.01 and 0.05 was performed using neutron powder diffraction at the LAD diffractometer at the ISIS spallation neutron source, Didcot, UK, and the D7 polarized neutron spectrometer at the Institut Laue-Langevin in Grenoble, France.   A powder diffraction pattern of \mbox{$\beta$--Mn$_{0.95}$In$_{0.05}$} is shown in Fig.~\ref{lad_diffraction}, together with a Rietveld refinement of the data, performed using the {\footnotesize GSAS} program~\cite{Larson94}.   Refinement of the D7 powder diffraction patterns was performed with the {\footnotesize FULLPROF}/{\footnotesize WinPLOTR} package~\cite{Carvajal93}.  Due to the significant neutron scattering length contrast between Mn and In ($b_{\rm{Mn}}=-3.73 $ fm, $b_{\rm{In}}=+4.06$ fm) the In site substitutional preference, and exact In concentration may be readily refined.  It was found that the In atoms - in common with Al and Sn - display a strong preference for the Mn2 sublattice, and therefore  the modeling of the data was performed with In substituents solely occupying the $12d$ Mn2 site. The refined fitting parameters obtained from the Rietveld refinements of the data for all samples studied are presented in table~\ref{tab1}.  Using the refined lattice constants and In concentrations, we have extracted the volume expansion as a function of In doping, shown in Fig.~\ref{vol_exp}.  Our data is in full agreement with that of Kohara~\cite{Kohara74}, indicating a volume expansion rate of $1.08 \pm 0.05$ \% per at. \% In.
\begin{table}
\caption{\label{tab1} Fitting parameters obtained by Rietveld refinement of the neutron powder diffraction pattern of \bmnin. $x_{\rm{nom}}$ and $x_{\rm{calc}}$ are the nominal and calculated In concentrations, $a$ is the cubic lattice constant, $d_1$ and $d_{2}$ are the Mn1 and Mn2 sublattice near neighbour distances, $R_{wp}$ is the weighted Bragg R-factor from the Rietveld refinement.}
\begin{ruledtabular}
\begin{tabular}{ccccccc}
$x_{\rm{nom}}$  &  $a$ (\AA) &  $d_1$ (\AA) & $d_{2}$ (\AA) &  $x_{\rm{calc}}$ & $R_{\rm{wp}}$\\
\hline
0\footnotemark[1]      & 6.320 & 2.367 & 2.646  & 0          & 3.95 & \\
0.01\footnotemark[2] & 6.340 & 2.363 & 2.647  & 0.014   & 2.91 & \\
0.05\footnotemark[1] & 6.433 & 2.401 & 2.692  & 0.050   & 6.04 & \\
0.05\footnotemark[2] & 6.444 & 2.401 & 2.694  & 0.056   & 3.32 & 
\end{tabular}
\end{ruledtabular}
\footnotetext[1]{Measured using the LAD diffractometer}
\footnotetext[2]{Measured using the D7 diffractometer}
\end{table}

\section{\label{xyz}Neutron polarization analysis of nuclear and magnetic short--range order}
Observations of the static magnetic ground--state structure of \bmnin alloys with $x=0.01$ and 0.05 were undertaken using the ``xyz'' neutron polarization analysis (NPA) technique, on the diffuse scattering spectrometer, D7, Institut Laue-Langevin, Grenoble~\cite{Stewart09}.  Using this technique, the magnetic structure factor, $S(Q)$, where $Q$ is the modulus of the neutron wavevector transfer, may be extracted from nuclear scattering and nuclear-spin-incoherent scattering contributions using a beam of spin-polarized neutrons, and analysing their final spin-states as a function of scattering angle, and polarization direction with respect to the scattering vector~\cite{Scharpf93}.  This technique is especially valuable in the case of disordered magnetic systems - where the magnetic scattering is distributed broadly in momentum-space, and is often weak and therefore difficult to separate from non-magnetic diffuse scattering contributions - such as strain or isotope disorder scattering.  Additionally, the use of NPA greatly facilitates the absolute calibration of the magnetic scattering cross--section in units of barns st$^{-1}$ atom$^{-1}$ which is proportional to the square of the total magnetic moment.  Therefore the extraction of absolute values of the diffuse magnetisation is straightforwardly achieved.

In the case of \bmnin, the analysis of the {\it nuclear} short--range order scattering will also provide valuable information on whether the In substituent atoms form a random solid solution with Mn on the Mn2 sublattice, or else whether they tend to either form clusters, or indeed {\it anti--clusters} ({\it i.e.} where atoms tend to separate themselves apart).   The magnetic diffuse scattering in turn, provides a direct measurement of the Mn spin--correlations, and therefore will determine the extent of any short--range magnetic order (magnetic clustering).  It is important to note that, since the impurity In atoms do not possess an atomic moment, any nuclear short--range order amongst the In atoms will likely have an effect on the extent of the Mn-Mn spin-correlations.

In the analysis of the NPA data, we make the tacit assumption that the magnetic scattering is ``static'' on a timescale set by the incident energy of the neutrons on D7.  In this case, a neutron wavelength of 3.1~\AA~was used, giving an incident neutron energy of \mbox{$E_i \simeq 8.5$ meV}.  In order to qualify as ``static'', any spin--fluctuations must have an energy of much less than this value, {\it i.e.} $\hbar \omega \ll E_i$.  This assumption is, to some extent, backed up by the muon spectroscopy results presented in section~\ref{muons}.

In a binary solid solution alloy with atomic species A and B, the diffuse scattering cross--section due to correlations between the binary species is given by~\cite{squires78},
\begin{eqnarray}
\label{nsro}
\left(\frac{d \sigma}{d \Omega}\right)_{\rm{diff}} 
	&=& \sum_{n=0}^{\infty} \left(\overline{b_ib_j}\right)_n 
		Z_n \exp\left(i\bm{Q}\cdot\bm{R}_n\right) - \left(\bar{b}\right)^2  \nonumber \\
	&=& x_A x_B \left(b_{\rm{A}}-b_{\rm{B}}\right)^2 \times \nonumber \\
	&  &  \sum_{n=0}^{\infty} Z_n\left(1-\frac{p_{\rm{A}}(n)}{x_A}\right) 
\exp\left(i\bm{Q}\cdot\bm{R}_n\right)
\end{eqnarray}
where $Z_n$ is the number of atoms, $p_{\rm{A}}(n)$ is the probability of finding an A atom in the $n^{\rm{th}}$ near neighbour shell around a central B atom and $x_A$ is the concentration of A atoms.  This equation was derived separately by Cowley~\cite{Cowley50b} and Warren~\cite{warren51}.  The term in brackets - which is the Fourier coefficient of the $n^{\rm{th}}$ near-neighbour shell term in the Fourier sum, is termed the {\it Warren-Cowley} short--range order parameter.
\begin{equation}
\alpha_n = 1 - \frac{p_{\rm{A}}(n)}{x_A} = 1-\frac{p_{\rm{B}}(n)}{x_B}
\end{equation}
where we have used the fact that the probabilities of finding an A atom in the $n^{\rm{th}}$ near neighbour shell around a central B atom, $p_{\rm{A}}(n)$, are related to the probabilities of finding a B atom in the $n^{\rm{th}}$ near neighbour shell around a central A atom, $p_{\rm{A}}(n)$, by the connecting relation, $x_A p_{\rm{B}}(n) = x_B p_{\rm{A}}(n)$. Warren-Cowley (WC) parameters have the following properties.  For the zero$^{\rm{th}}$ near neighbour shell, the probability $p_{\rm{A}}(0) = 0$ and therefore, $\alpha(0) = 1$. For a random alloy, the probability $p_{\rm{B}}$ (cf. of finding a B atom around an A atom) is equal to the concentration of B atoms, $x_B$. Therefore, $\alpha_{n > 0} = 0$, and Eq.~\ref{nsro} reduces to the expression for  Laue monotonic scattering, $x_A x_B \left(b_{\rm{A}} - b_{\rm{B}}\right)^2$. The limiting values of the WC parameters are $\alpha_{\rm{max}} = 1$; representing  a central atom entirely surrounded by like atoms and $\alpha_{\rm{min}} = 1-1/x_A=-x_B/(1-x_B)$; for a central atom entirely surrounded by unlike atoms. For the analysis of the polycrystalline samples used in this study, we take the orientational average over all directions of the position vector $\bm{R}_n$, and the expression for the nuclear diffuse scattering becomes
\begin{eqnarray}
\label{pownsro}
\left(\frac{d \sigma}{d \Omega}\right)_{\rm{diff}} = x_A x_B (b_{\rm{A}}-b_{\rm{B}})^2 \times \nonumber \\
\left[1+\sum_{n=1}^{\infty} Z_n \alpha_n\frac{\sin QR_n}{QR_n}\right]
\end{eqnarray}

\begin{figure}
\includegraphics[width=3.2in]{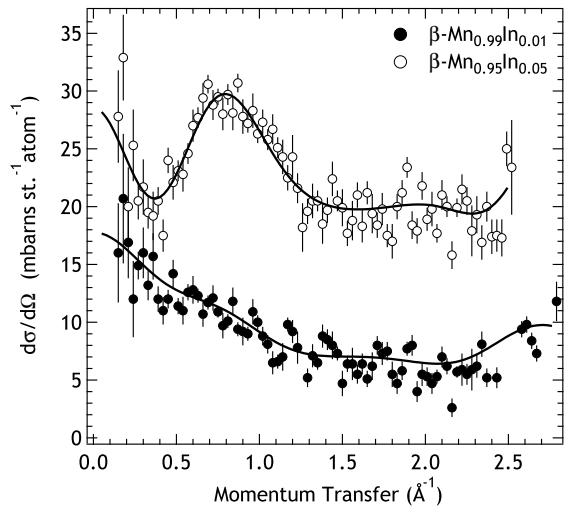}
\caption{\label{nucsro}The nuclear diffuse scattering cross--section of \bmnin with $x=0.01$ (solid circles) and 0.05 (open circles).  The nuclear Bragg peaks have been removed from the data.  The solid lines are fits to the diffuse nuclear scattering using a Warren-Cowley formalism, achieved using our RMC fitting procedure.}
\end{figure}
\begin{figure}
\includegraphics[width=3.2in]{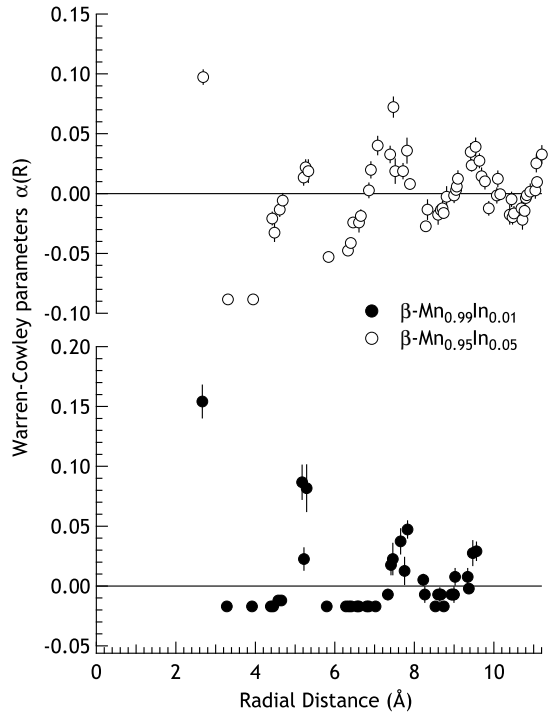}
\caption{\label{nuccorrs}The Warren-Cowley (WC) parameters extracted from the RMC model fits to the nuclear diffuse scattering data shown in Fig.~\ref{nucsro} for the first forty near neighbour shells of $\beta$--Mn$_{0.99}$In$_{0.01}$ (solid circles) and the first sixty near neighbour shells of $\beta$--Mn$_{0.95}$In$_{0.05}$ (open circles).}
\end{figure}

The nuclear diffuse scattering ({\it i.e.} with Bragg peaks removed) measured on D7 for \bmnin with $x=0.01$ and $x=0.05$ is shown in Fig.~\ref{nucsro}.  The data has been modeled according to Eq.~\ref{pownsro} using a reverse monte-carlo (RMC) procedure developed by us, and described in Ref.~\onlinecite{Stewart08}.  A model of $8 \times 8 \times 8$ unit cells - with only the Mn2 positions present - was used as input to the model.  Indium atoms were added to the model at initially random positions.  From the generated In-doped lattice, the WC parameters as a function of radial distance are found and the diffuse cross--section calculated using Eq.~\ref{pownsro}.  This calculated cross-section is then compared with the data, using a standard $\chi^2$ goodness-of-fit parameter.  In atoms may then be randomly swapped with Mn atoms, and the cross-section recalculated.  Swaps which result in a reduced  $\chi^2$ are accepted, and those which increase  $\chi^2$ are rejected.  The procedure continues until the point at which $\chi^2$ stops decreasing.  In order to verify that the final calculation is independent of initial conditions, the RMC procedure is run several times, and the results averaged. The fitted cross--section is shown as solid lines in Fig.~\ref{nucsro}, and the WC parameters extracted from the fitted model are plotted in Fig.~\ref{nuccorrs}.  The fitted Laue monotonic cross--sections were found to be $8 \pm 3$ mbarns st.$^{-1}$ atom$^{-1}$ for $\beta$--Mn$_{0.99}$In$_{0.01}$ and $22 \pm 9$ mbarns st.$^{-1}$ atom$^{-1}$ for $\beta$--Mn$_{0.95}$In$_{0.05}$.  These values are in good agreement with those calculated using $x_A x_B \left(b_{\rm{A}} - b_{\rm{B}}\right)^2$ (6 mbarns st.$^{-1}$ atom$^{-1}$ and 29 mbarns st.$^{-1}$ atom$^{-1}$ respectively), proving that the absolute normalization of the neutron data is reliable.  Interestingly, the In atoms show a tendency to cluster together with a strongly positive WC parameter in the first near neighbour shell of both the $x=0.01$ and $x=0.05$ \bmnin samples.  This is unlike the case of Al impurities in $\beta$--Mn where a tendency to anti--cluster was observed~\cite{Stewart08}.  However, the clusters of In atoms are small  - 2 to 3 atom clusters extending over a correlation length of the order of 2.5 \AA, with an average cluster separation of $\sim 2.5$ \AA~for $\beta$--Mn$_{0.99}$In$_{0.01}$ and $\sim 2.0$ \AA~for $\beta$--Mn$_{0.95}$In$_{0.05}$.

\begin{figure}
\includegraphics[width=3.2in]{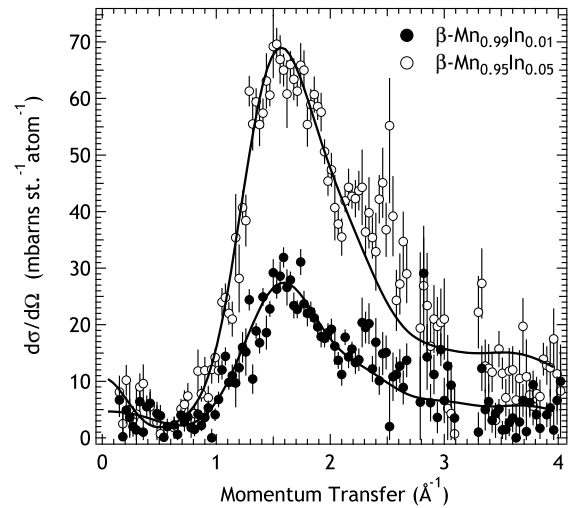}
\caption{\label{magsro}The magnetic diffuse scattering cross--section of \bmnin\ with $x=0.01$ (solid circles) and 0.05 (open circles).   }
\end{figure}

Having modeled the nuclear disorder scattering, one can then proceed to interpretation of the magnetic scattering cross--section.   This is shown in Fig.~\ref{magsro} and is seen to consist of a broad peak in momentum space, centred on a neutron momentum transfer of $Q \simeq 1.6$ \AA $^{-1}$.  The integrated magnetic cross--section increases dramatically on increasing In concentration, peaking at $\sim 25$ mbarns st.$^{-1}$ atom$^{-1}$ for $\beta$--Mn$_{0.99}$In$_{0.01}$ and $\sim 70$ mbarns st.$^{-1}$ atom$^{-1}$ for $\beta$--Mn$_{0.95}$In$_{0.05}$.  This may be due to either an increasing Mn moment on introduction of In, or it may be due to a slowing down of magnetic fluctuations on increasing In doping, resulting in an increased contribution to the ``static'' magnetic cross--section measured on D7, as discussed above.  

In the case of a disordered (paramagnetic or frozen) magnetic configuration, the magnetic neutron scattering differential cross--section is given by the standard expression,
\begin{eqnarray}
\label{magsro_eq}
\left(\frac{d\sigma}{d\Omega}\right)_{\rm{mag}} = \frac{2}{3}\left(\frac{\gamma_nr_0}{2}\right)^2 f^2(\vec{Q}) g_S^2 S(S+1) \nonumber \\
\times \left[1 + \sum_{n=1}^{\infty} Z_n\frac{\langle\bm{S}_0 \cdot \bm{S}_n\rangle}{S(S+1)}\frac{\sin QR_n}{QR_n}\right]
\end{eqnarray}
whereby the average spin--spin correlation functions $\langle\vec S_0 \cdot \vec S_n\rangle$ for the $n$th near neighbour radial shell around a central Mn atom at the origin may be computed.   $\gamma_n$ is the neutron gyromagnetic ratio, $r_0$ is the classical electron radius, $r_0 = \mu_0e^2/4\pi m_e$, $f(Q)$ is the magnetic form factor of the magnetic species used in the model (Mn$^{3+}$ in this case), $g_s^2S(S+1)$ is the squared magnetic moment of the magnetic species and $Z_n$ and $R_n$ are the co-ordination number and radial distance of the $n$th near neighbour shell respectively.  Grouping the constants together, the pre--factor $2/3\left(\gamma_n r_0/2\right)^2 = 0.049$ barns.  Therefore for a paramagnetic system, with zero average spin-spin correlations, the absolute squared magnetic moment per atom is straightforwardly calculated from Eq.~\ref{magsro_eq}.

The turning points of Eq.~\ref{magsro_eq} are such that the the position of the peak in the magnetic cross--section at $Q_{\rm{max}}$ is related to the antiferromagnetic (negative turning point) correlation distance as $R_{\rm{AF}} \simeq 4.5 / Q_{\rm{max}}$.  This naive argument leads to an estimate of $R_{\rm{AF}} \simeq 2.8$~\AA\ which is comparable to the near-neighbour distance between Mn2 sites of between 2.65 and 2.7~\AA (see Tab.~\ref{tab1}), but somewhat larger than the Mn1 near neighbour distance of 2.36~\AA, lending credence to the contention that the Mn2 sublattice is dominantly magnetic.

In order to go a step beyond this simple analysis, we have performed a RMC fit of the data, analogous to that used to model the nuclear diffuse scattering.  As input to the RMC calculation, we employed the atomic model generated by the nuclear fitting procedure.  Heisenberg spins were then assigned to the Mn atoms in the model, and zero spin on the In sites.  In this way, any influence of In concentration fluctuations on the Mn spin--spin correlation functions is automatically taken into account.  In each RMC step, Mn spins are randomly chosen and then rotated by a random angle.  The magnetic cross--section is then calculated using Eq.~\ref{magsro_eq} and compared to the data.   The fitted squared magnetic moment per atom (within the experimental time--window) was found to be $0.5 \pm 0.1~\mu_{\rm{B}}^2$ for $\beta$--Mn$_{0.99}$In$_{0.01}$ and $1.2 \pm 0.2~\mu_{\rm{B}}^2$ for $\beta$--Mn$_{0.95}$In$_{0.05}$; representing either a significant slowing down of the Mn spin--fluctuations, or an enhancement of the Mn magnetic moment on In doping. 

\begin{figure}
\includegraphics[width=3.2in]{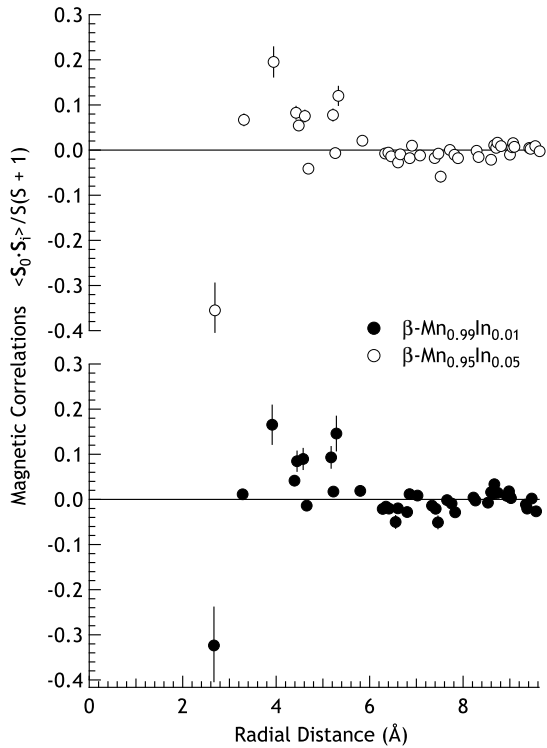}
\caption{\label{magcorrs}The normalised magnetic spin-spin correlations extracted from the RMC model fits to the magnetic diffuse scattering data shown in Fig.~\ref{magsro} for the first forty near neighbour shells of $\beta$--Mn$_{0.99}$In$_{0.01}$ (solid circles) and $\beta$--Mn$_{0.95}$In$_{0.05}$ (open circles).}
\end{figure}

The magnetic correlations extracted from the RMC model, plotted in Fig.~\ref{magcorrs}, show dominant first near neighbour antiferromagnetic correlations at $R \simeq 2.7$ \AA\, as hinted at by our naive inversion of the peak in $S(Q)$.  Significant magnetic correlations continue till $\sim 5.4$ \AA\ - the tenth near neighbour shell, with correlation terms close to zero thereafter.   The detailed $R-$dependence of the spin--spin correlations in the $x=0.01$ and $x=0.05$ samples is quantitatively identical, with two shells exhibiting strong positive correlations at $R \simeq 4$~\AA\ (third neighbour shell) and $R \simeq 5.4$~\AA\ (tenth neighbour shell), leading us to conclude that the nature of the short-range Mn spin configurational structure is \emph{unaffected} by the level of In concentration.   Indeed these peaks in the correlation spectrum are exactly the same as those seen in $\beta$--Mn(Co) alloys, in which the Co substituents reside on the Mn1 sublattice alone.  We therefore claim that the disruption of the magnetic Mn2 sublattice by the presence of non-magnetic In atoms has minimal effect on the ground--state magnetic structure of \bmn.    The spin--spin correlations are of somewhat shorter range than was previously found for both $\beta$--Mn(Al)~\cite{Stewart08} and $\beta$--Mn(Co)~\cite{Stewart09b} dilute alloys.

\section{\label{muons}Muon Spin Relaxation measurements}
Muon spin relaxation ($\mu$SR) has, in the past, proved to be a very valuable probe of dynamical properties of spin--fluctuating itinerant magnets, such as YMn$_2$~\cite{Cywinski94,Rainford95,Cywinski91}.  Indeed, previous $\mu$SR studies of \bmn\ alloys with Al and Ru dopants have revealed magnetic dynamical phase transitions inaccessible to other techniques.~\cite{Stewart99,Leavey07}  We have therefore performed a $\mu$SR study of \bmnin\ in order to characterise the spin--dynamics of the system, and to examine the apparent formation of a spin--glass state at low temperatures

\begin{figure}
\includegraphics[width=3.2in]{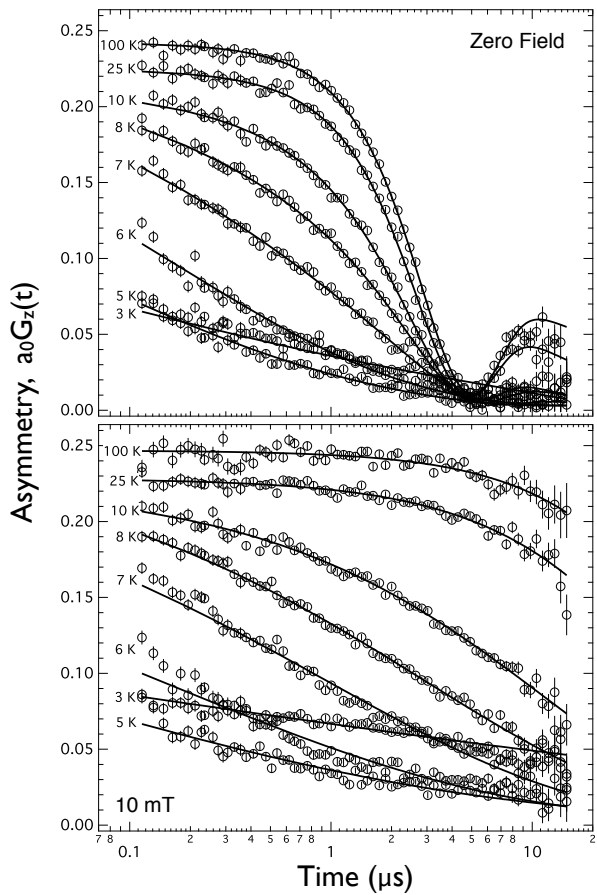}
\caption{\label{muonspectra}The asymmetry spectra measured on the MuSR spectrometer for $\beta$--Mn$_{0.99}$In$_{0.01}$ as a function of temperature.  The top panel shows the spectra measured in zero field (ZF) with the solid lines representing fits to Eq.~\ref{zffits}.  The lower panel shows the data taken in 10 mT longitudinal field (LF), with the solid lines showing fits to Eq.~\ref{lffits}}
\end{figure}

$\mu$SR spectra of \bmnin\ with $x=$0.01 and 0.05 were measured using the MuSR spectrometer at the ISIS pulsed muon facility, UK.   The decay-positron asymmetry function was measured as a function of time, in both zero field (ZF) and a longitudinal field (LF) of 10 mT.  The asymmetry functions measured for $\beta$--Mn$_{0.99}$In$_{0.01}$ as a function of temperature are shown in Fig.~\ref{muonspectra}.  

The ZF spectra (Fig.~\ref{muonspectra} - top panel) were modeled using 
\begin{equation}
\label{zffits}
	a_0 G_z (t) = a_1 G_{\rm{KT}}(t) \times G_{\rm{Mag}}(t) + a_{\rm{bg}}
\end{equation}
where $a_0$, $a_1$ and $a_{\rm{bg}}$ are the initial, relaxing and background initial asymmetries respectively, and $G_{\rm{KT}}(t)$ is the static zero-field Kubo-Toyabe function,
\begin{equation}
\label{GKT}
	G_{\rm{KT}}(t) = \frac{1}{3} + \frac{2}{3}\left(1-\sigma^2 t^2\right) \exp\left(-\frac{\sigma^2 t^2}{2}\right)
\end{equation}
where the nuclear depolarization rate $\sigma^2 = \gamma_{\mu}^2 \Delta^2$, where $\gamma_{\mu}$ is the muon gyromagnetic ratio ($= 8.52 \times 10^8$ rad. s$^{-1}$ T$^{-1}$) and $\Delta^2$ is the second moment of the static nuclear field distribution.  The magnetic part of the relaxation function, 
\begin{equation}
\label{gmag}
	G_{\rm{Mag}} = \exp\left[-\left(\lambda t\right)^{\beta}\right]
\end{equation}
represents the relaxation contribution from the dynamic magnetic fields associated with fluctuating atomic spins.  This particular relaxation function is known as a \emph{stretched exponential} function, and has proved very successful in the description of the dynamics of spin--glasses in the paramagnetic regime.~\cite{Campbell94,Campbell99} The multiplicative combination for the magnetic and nuclear relaxation functions in Eq.~\ref{zffits} is valid providing that the nuclear and atomic fields contribute independently to the muon depolarization.  

\begin{figure}
\includegraphics[width=3.2in]{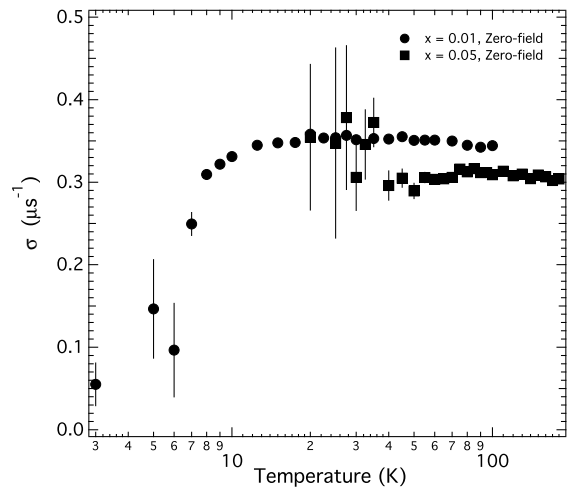}
\caption{\label{sigma}The temperature dependence of the nuclear depolarization rate, $\sigma$, for \bmnin\ with $x=$ 0.01 (circles) and 0.05 (squares).}
\end{figure}

Fig.~\ref{sigma} shows the temperature dependence of the nuclear depolarization rate, $\sigma$, for the $x=$ 0.01 and 0.05 compositions.  $\sigma$ is constant down to around 7 K for the $x=0.01$ sample and down to the lowest temperatures measured in the case of $x=$ 0.05.  For this latter sample, measurements were taken down to 20 K, being the temperature at which the muon asymmetry spectra are effectively depolarized, indicating frozen order.  The high--temperature values are found to be $\sigma = 0.350 \pm 0.001$ MHz for the $x=$ 0.01 sample,  and  $\sigma = 0.308 \pm 0.001$ MHz for the $x=$ 0.05 sample.  The temperature at which $\sigma$ is seen to deviate from the high-temperature value marks the point at which the slowing atomic spins might start to couple to the nuclear magnetism, resulting in a so--called ``double--relaxation'' process.~\cite{Kadono90}   This then demarcates the limit of validity of the independent relaxation channel model of Eq.~\ref{zffits}.  

A standard method of suppressing the nuclear contribution to the muon asymmetry function is by application of a small field, in a direction parallel to the muon polarization (longitudinal).  This results in the muon being effectively de-coupled from the much weaker static nuclear dipolar fields at the interstitial muon site in the material.   We have, accordingly, also performed longitudinal field (LF) measurements of the muon asymmetry spectra in a field of 10 mT.  This field was found to be the minimum required to fully de-couple the nuclear relaxation channel.  The LF muon spectra for the $x=$ 0.01 sample are shown in the lower panel of Fig.~\ref{muonspectra}.   With the nuclear channel suppressed, the muon asymmetry function may now be modeled by
\begin{equation}
\label{lffits}
	a_0 G_z (t) = a_1 G_{\rm{Mag}}(t) + a_{\rm{bg}}
\end{equation}
with $ G_{\rm{Mag}}(t)$ given by Eq.~\ref{gmag}.  The LF muon spectra and fits are plotted in the lower panel of Fig.~\ref{muonspectra}.   Application of a field might also be expected to have some effect on the magnetic part of the muon relaxation, $G_{\rm{Mag}}$, depending on how strong the interstitial local fields are at the muon site, and on the field-fluctuation rate.  In general for spin-glasses, the magnetic relaxation scales with field as
\begin{equation}
	G_{\rm{Mag}}(H,t) = G_{\rm{Mag}}(t/H^{\gamma})
\end{equation}
where the exponent $\gamma$ can give information as to the nature of the spin--spin autocorrelation function in the material~\cite {Keren96}.  For many spin-glasses this autocorrelation function is found to follow a cutoff power--law of the form 
\begin{equation} 
\label{autocorr}
\langle \bm{S}_i(t) \cdot \bm{S}_i(0) \rangle \propto t^{-\alpha} \exp\left[ -\left(\lambda_h t\right)^{\beta_h}\right] 
\end{equation}
where the fluctuation rate, $\lambda_h$ and the exponent, $\beta_h$ have been given the subscript $h$ to indicate that they are different to the muon relaxation rate and exponent in Eq.~\ref{gmag}.   Such non-exponential relaxation emerges directly from Ogielski's~\cite{Ogielski85} Monte Carlo calculations of spin glasses although more recently a fundamental thermodynamic origin for such a relaxation function in spin glasses has  been suggested:   Pickup et al ~\cite{Pickup09} explicitly considered macroscopic interactions and hierarchically constrained dynamics in spin glasses, and showed the associated interaction parameter can be directly related to a normalized Tsallis~\cite{Tsallis95} non-extensive entropy parameter which in turn is found to exhibit a universal scaling with reduced temperature for a number of spin glasses.  Using this autocorrelation function (Eq.~\ref{autocorr}), Keren was able to model the expected reduction in the muon relaxation rate as a function of external field, and in doing so, confirmed that the Ogielski function was appropriate for the spin--glass AgMn.  He also noted, in common with the previous work of Campbell and co-workers,~\cite{Campbell94} that the exponent $\beta$ in $G_{\rm{Mag}}$ tends to a value of 1/3 as the temperature approaches the spin--glass freezing temperature, $T_g$.  

\begin{figure}
\includegraphics[width=3.2in]{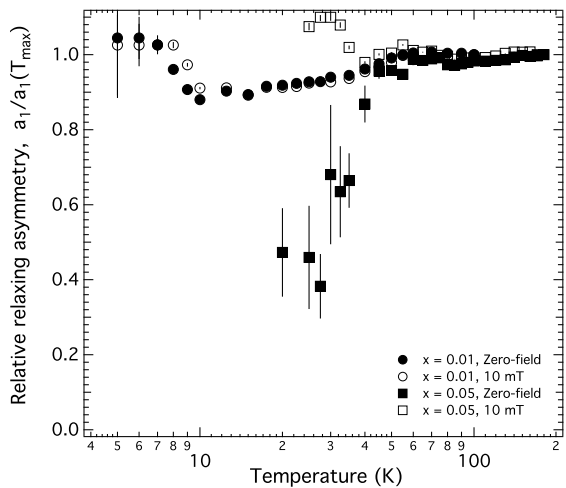}
\caption{\label{a1}The temperature dependence of the relaxing part of the initial asymmetry, $a_1$, normalised to the highest temperature value, for \bmnin\ with $x=$ 0.01 (circles) and 0.05 (squares) in zero field (solid symbols) and 10 mT (open symbols).  }
\end{figure}

\begin{figure}
\includegraphics[width=3.2in]{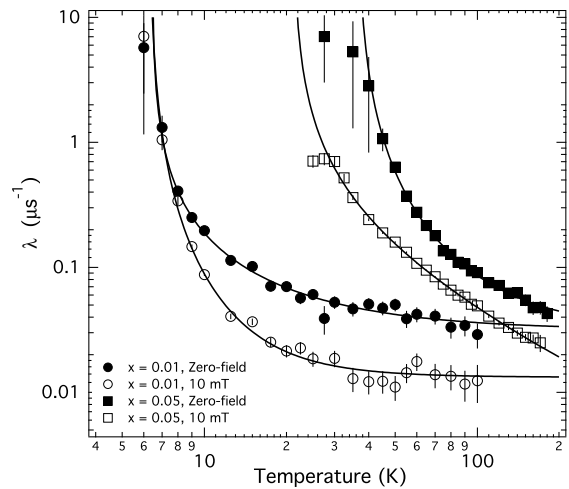}
\caption{\label{lambda}The temperature dependence of the muon relaxation rate, $\lambda$,  for \bmnin\ with $x=$ 0.01 (circles) and 0.05 (squares) in zero field (solid symbols) and 10 mT (open symbols).  The solid lines show fits to a critical form given by Eq.~\ref{critical}}. 
\end{figure}

\begin{figure}
\includegraphics[width=3.2in]{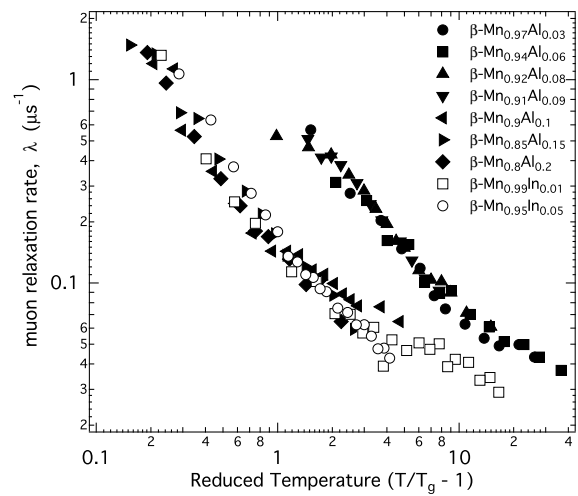}
\caption{\label{scaling}The zero field muon relaxation rate, $\lambda$, scaled to the spin--glass freezing temperature $T_g$  for \bmnin\ with $x=$ 0.01 (hollow squares) and 0.05 (hollow circles).  The data for \bmnal\ with $0.03 \le x \le 0.2$ previously presented in Ref.~\onlinecite{Cywinski04} are also shown as solid symbols.}. 
\end{figure}

Fig.~\ref{a1} shows the fitted initial relaxing asymmetry, $a_1$, of \bmnin\ normalised to high temperature values.  For the $x = 0.01$ concentration, $a_1$ is nearly independent of temperature.  There is a slight reduction from $\sim$ 1 to $\sim$ 0.9 between 50 K and 10 K.  Below 10 K the initial asymmetry recovers to $\sim$ 1.  For the $x=0.05$ concentration, $a_1$ deviates from $\sim$ 1 at around 40 K.  And - in the zero-field case - approaches around 1/3 at around 30 K.  This behaviour of a drop in the initial asymmetry indicates a transition to a static magnetic state - with immediate depolarization of the 2/3 components of muon polarization transverse to the static atomic fields (randomly oriented).
\begin{figure}
\includegraphics[width=3.2in]{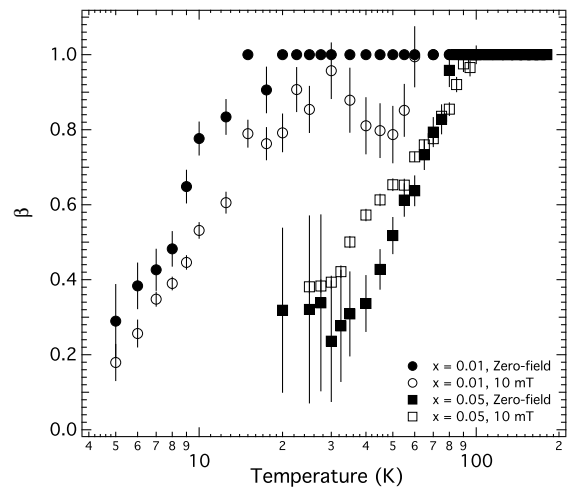}
\caption{\label{beta}The temperature dependence of the muon relaxation exponent, $\beta$, for \bmnin\ with $x=$ 0.01 (circles) and 0.05 (squares) in zero field (solid symbols) and 10 mT (open symbols).  }
\end{figure}
Further evidence for a dynamical transition in the \bmnin\ alloys studied comes from the temperature dependence of the muon exponent $\beta$, from Eq.~\ref{gmag}, plotted in Fig.~\ref{beta}.   Both the ZF and 10 mT runs show a qualitatively similar behaviour, with $\beta$ decreasing from 1 at high temperatures, to around 1/3 at $T \sim $ 5 K for the $x=0.01$ composition, and $T \sim 30 $~K for the $x=0.05$ composition.   The temperature dependence of the muon relaxation rate, $\lambda$, is shown in Fig.~\ref{lambda}.
It has been possible to fit the muon relaxation rate to a critical form
\begin{equation}
\label{critical}
	\lambda = \lambda_0 \left(\frac{T_g}{T-T_g}\right)^\zeta + \lambda_{\rm{bg}}
\end{equation}
in order to gain an accurate value for the freezing temperature.   $\lambda_{\rm{bg}}$ is a non--temperature--dependent contribution to the relaxation rate.   This background contribution was found to be 0.03 $\mu$s$^{-1}$ for the ZF runs and 0.01 $\mu$s$^{-1}$ for the 10 mT runs.   The ZF freezing temperatures found were; $T_g = 6.4 \pm 0.6$ K for $x=0.01$ and  $T_g = 35 \pm 2$ K for $x=0.05$.   The 10~mT freezing temperatures found were; $T_g = 6.2 \pm 0.2$ K for $x=0.01$ and  $T_g = 21 \pm 2$ K for $x=0.05$.    The values of $T_g$ found for the $x=0.01$ composition are similar to that found using DC susceptibility, presented in Sec.~\ref{prep}, of $T_g \sim 5$ K.  However there is a clear discrepancy between the freezing temperatures deduced for the $x=0.05$ composition, with ZF $\mu$SR, LF $\mu$SR and DC susceptibility all giving very different values (35 K, 21 K and 40 K respectively).    From the plot of the temperature dependence of the initial asymmetry, $a_1$ (Fig.~\ref{a1}) it is clear that the ZF and LF runs of the $x=0.05$ sample present very different behaviours, with clear loss of initial asymmetry at low temperatures in the ZF case, and no loss (even apparent increase) in the initial asymmetry in the LF case.   Clearly, the application of a field - even as small as 10 mT - has a measurable effect of the magnetic fluctuations in \bmnin\ and therefore, the ZF runs should be regarded as a more reliable indicator of the magnetic ground state.   The scaling exponent found for each of the alloys was found to be in the range $ 1.1 \lesssim \zeta \lesssim 1.5$, allowing us to plot $\lambda$ on a universal scaling plot vs. reduced temperature $T/T_g - 1$, in Fig.~\ref{scaling}.
Previously we noted that such a scaling plot revealed a clear difference in behaviour between \bmnal\ alloys with Al concentrations $x \le 0.09$ and those with $x > 0.1$~\cite{Cywinski04}.  This difference was associated with a quantum spin--liquid to spin--glass transition at around 9 at.\% Al, with a concomitant shift from simple exponential to stretched exponential muon depolarization.  We note from Fig.~\ref{scaling} that the ZF muon relaxation rates show qualitatively the same scaling behaviour as the concentrated Al alloys, falling onto the same scaling regime as the highly concentrated spin--glass \bmnal\ alloys.  

There is also evidence for residual fluctuations below the freezing temperatures, by inspection of the low temperature asymmetry functions.  If the local fields at the muon site were completely static, one would expect a flat, non-depolarizing muon spectrum, at a level of 1/3 of the high temperature initial asymmetry (since the spin system is randomly oriented).  In fact, there is considerable depolarization below the freezing transitions in both \bmnin\ compositions studied (see Fig.~\ref{muonspectra}).  It is possible that \bmnin\ may become increasingly static as the temperature is further decreased below the supposed freezing transition temperature, however, these temperatures were not accessed during the measurement.  

\section{\label{disc}Discussion}
The development of a ``static'' magnetic ground state in a dilute alloys of \bmn\ is almost guaranteed, regardless of the choice of substituent.  Studies of \bmn\ alloyed with Al, Sn, In, Zn, Fe, Co, Ni, Ru, Ir, and Os all show the development of a static - but in most cases, disordered - magnetic ground state.  In this study, we have concentrated on In doped samples in order to gain insight into the exact mechanism by which the magnetic fluctuations in the spin--liquid ground state of pure \bmn\ are quenched.  

From DC susceptibility, we have found the characteristic signature of a spin--glass like susceptibility in \bmnin\, with a cusp at finite temperature, and bifurcation of the FC and ZFC branches indicative of the history--dependence expected for a glassy magnet with broken ergodicity.  This cusp appears at the lowest indium concentrations studied (1 at.\% In), and exhibits a roughly $T$-linear behaviour as a function of doping.  The $\mu$SR studies confirm the existence of a dynamical transition, typical of a spin--glass, for \bmnin\ with $x \ge 0.01$.   There is also evidence for residual slow dynamics below the deduced freezing temperatures.   Despite this dramatic change in the susceptibility ({\it N.B.} bulk susceptibility at wavevector $Q=0$), and the creation of a static magnetic ground--state, a detailed investigation of the static \emph{Q-dependent} susceptibility using polarized neutron scattering has shown that there is actually very little difference between the $Q$-dependences of the magnetic structure factor $S(Q)$ measured for the $x=0.01$ and 0.05 compositions of \bmnin\ - other than an overall increase in intensity.   The measured magnetic structure factors (Fig.~\ref{magsro}) highlight the fact that the bulk DC susceptibility samples only a very insignificant part of the $Q$-dependent susceptibility in \bmnin.   Since the RMC analysis reveals that the real-space spin--correlations are almost identical between the two compositions, we can state that the spin--liquid to spin--glass transition is not driven by a relief of the frustrated Mn--Mn exchange couplings on the Mn2 sublattice which would result in longer range spin--correlations.  It appears instead to be entirely due to the massive volume expansion of the lattice, resulting in more localised Mn moments, and a strong damping of longitudinal spin--fluctuations.   This is reminiscent of the spin--liquid to spin--glass dynamical transition observed in \bmnal\ in which a sharp cross-over from a dynamic to a static ground--state was observed at an Al doping level of 9 at. \%.~\cite{Stewart99}  Here the creation of a static response is achieved immediately at 1 at. \% In.  We note that the lattice constant of $\beta$--Mn$_{0.99}$In$_{0.01}$ of 6.34 \AA\ is the same as that of $\beta$--Mn$_{0.92}$Al$_{0.08}$,~\cite{Stewart08} lending further evidence supporting the assertion that localization of the Mn spins on expansion of the lattice is the reason for the development of a static ground--state, and not disruption of the Mn2 sublattice by non--magnetic impurities.  

It is also instructive to compare these results to those obtained for $\beta$--Mn(Co) alloys~\cite{Stewart09b}.  In that case the Co atoms reside on the Mn1 sublattice, and the lattice constant is almost independent of Co concentration.  The spin--correlations in $\beta$--Mn(Co) are of much longer range than those observed in this study, or in a similar study of $\beta$--Mn(Al).~\cite{Stewart08}.   In the case of $\beta$--Mn(Co), the data indicated the presence of a moment on the Mn1 sublattice.  Here, there is no such indication - with the spin--correlations being well described by a model involving only the Mn2 sublattice.  An attempt to model the data using both Mn sub-lattices did not result in an improvement of the fit to the polarized neutron diffraction data.  In the case of $\beta$--Mn(Ru), the Ru atoms are also sited on the Mn1 sublattice, but the rate of volume expansion of the lattice is very marked, at a rate of $\sim 0.4$ \% per at.\% Ru.~\cite{Sasao01}  In that case, not only is there the stabilization of long-range magnetic order, but there is evidence supporting the existence of localized moments on both Mn1 and Mn2 sublattices.   Thus we conjecture that the criteria for the formation of a unique long--range ordered magnetic structure in \bmn\ are a considerable increase in the lattice constant and an \emph{undisrupted} Mn2 sublattice.

\section{\label{conc}Conclusions}
We have carried out a full survey of dilute indium doped \bmn\ alloys in order to characterise the influence of atomic disorder and lattice expansion on the low--temperature magnetic properties.  Initial characterization using DC susceptibility shows the development of a spin--glass state for indium concentrations $> 1$ at. \%.  Neutron scattering investigations reveal that the indium impurities reside on the ``magnetic'' Mn2 sublattice, and show a slight tendency to cluster together.  There is a massive expansion of the \bmn\ lattice on indium doping of around 1.1 \% per at.\% In.  Despite this, the short-range magnetic correlations measured using neutron polarization analysis are independent of indium concentration.  We conclude therefore that the Mn--Mn spin correlations are unaffected by doping the Mn2 sublattice with non-magnetic indium, and hence that the geometrical frustration associated with the lattice is not lifted.   Using muon spin relaxation we have confirmed the existence of a static spin--glass ground--state, but with residual slow spin--dynamics below the freezing transition.  

\acknowledgments
The authors acknowledge very helpful discussions with S. Giblin, and for assistance with the susceptibility measurements.  We thank the ILL for supporting the PhD studentship of JMH.  We are also grateful for experimental support both at the ILL and ISIS neutron and muon facilities.

\bibliography{allrefs}
\end{document}